\documentclass[twocolumn,showpacs,amsmath,amssymb,prd]{revtex4}
\newcommand{\beq}{\begin{equation}}
\newcommand{\eeq}{\end{equation}}
\newcommand{\beqray}{\begin{eqnarray}}
\newcommand{\eeqray}{\end{eqnarray}}
\def\grad{\vec\nabla}
\def\div{\vec \nabla\cdot}
\def\gf{\grad\phi}
\def\gfs{(\grad\phi)^2}
\def\sss{\scriptscriptstyle}
\def\^#1{^{\sss #1}}
\def\_#1{_{\sss #1}}
\def\gmn{g_{\sss\mu \nu}}
\def\Gmn{g^{\sss\mu \nu}}

\def\oot{{1\over 2}}
\def\a{\alpha}

\def\c{\gamma}
\def\d{\delta}
\def\m{\mu}
\def\n{\nu}
\def\r{\rho}
\def\l{\lambda}

\def\deriv#1#2{{d#1\over d#2}}

\def\der#1{{}_{\sss,#1}}

\def\cd#1{{}_{\sss;#1}}

\def\pd#1#2{{\partial#1\over \partial#2}}

\def\div{\vec{\hbox{\mqq\char'162}}\cdot}
\def\grad{\vec{\hbox{\mqq\char'162}}}

\def\div{\vec\nabla\cdot}
\def\grad{\vec\nabla}

\def\dDx{d^{D+1}x}

\def\OD{\Omega_D}
\def\ODo{\Omega_{D-1}}
\def\gh{(-g)^{1/2}}
\def\gmh{(-g)^{-1/2}}

\def\l0{\ell_{0}}

\def\rar{\rightarrow}

\def\l{\lambda}
\def\t{\theta}

\def\f{\phi}
\def\z{\zeta}
\def\c{\gamma}

\def\r{\rho}
\def\m{\mu}
\def\dt{\der{t}}

\def\d{\delta}
\def\heta{\hat\eta}
\def\vv{\vec v}
\def\a{\alpha}

\def\vr{\vec r}
\def\vR{\vec R}

\def\vu{\vec u}
\def\vU{\vec U}

\def\drD{d^Dr}
\def\dRD{d^DR}
\def\rs{\vr_*}
\def\hs{\hat s}

\def\hpsi{\hat \psi}
\def\vvs{\vv_*}

\begin{document}

\title{Massive particles in acoustic space-times\\
 emergent inertia and passive gravity}
\author{Mordehai Milgrom }
\affiliation{ The Weizmann Institute Center for Astrophysics}

\begin{abstract}
 I show that massive-particle dynamics can be
simulated by a weak, external perturbation on a potential flow in
an ideal fluid. The perturbation defining a particle is dictated
in a small (spherical) region that is otherwise free to roam in
the fluid. Here I take it as an external potential that couples to
the fluid density or as a rigid distribution of sources with
vanishing total out-flux. The effective Lagrangian for such
particles is shown to be of the form $ mc^2\ell(U^2/c^2)$, where
$\vU$ is the velocity of the particle relative to the fluid and
$c$ the speed of sound. This can serve as a model for emergent
relativistic inertia a la Mach's principle with $m$ playing the
role of inertial mass, and also of analog gravity where $m$ is
also the passive gravitational mass. The mass $m$ depends on the
particle type and intrinsic structure (and on position if the
background density is not constant), while $\ell$ is universal:
For $D$ dimensional particles $\ell\propto F(1,1/2;D/2;U^2/c^2)$
($F$ is the hypergeometric function). These particles have the
following interesting dynamics: Particles fall in the same way in
the analog gravitational field mimicked by the flow, independent
of their internal structure, thus satisfying the weak equivalence
principle. For $D\le 5$ they all have a relativistic limit with
the acquired energy and momentum diverging as $U\rar c$. For $D\le
7$ the null geodesics of the standard acoustic metric solve our
equation of motion. Interestingly, for $D=4$ the dynamics is very
nearly Lorentzian: $\ell\propto -mc^2\c^{-1}\l(\c)$ (up to a
constant), with $\l=(1+\c^{-1})^{-1}$ varying between 1/2 to 1
($\c$ is the ``Lorentz factor'' for the particle velocity relative
to the fluid). The particles can be said to follow the geodesics
of a generalized acoustic metric of a Finslerian type that shares
the null geodesics with the standard acoustic metric. In vortex
geometries, the ergosphere is automatically the static limit. As
in the real world, in ``black hole'' geometries circular orbits do
not exist below a certain radius that occurs outside the horizon.
There is a natural definition of antiparticles; and I describe a
mock particle vacuum in whose context one can discuss, e.g.,
particle Hawking radiation near event horizons.

\end{abstract}

\maketitle

 \section{\label{seci}{Introduction}}

It is well documented that the propagation of acoustic waves in
inviscid, barotropic, irrotational background flows bears some
enlightening resemblances to propagation of light in curved space
times (see the seminal paper of Unruh\cite{unruh81} and many
subsequent expansions; e.g., \cite{jacobson91, visser93}, and the
recent extensive review in \cite{barcelo05}): The flow potential,
$\eta$, describing weak acoustic waves moving on a given
background $D$-dimensional flow satisfies the wave equation
 \beqray \Box\eta\equiv
 (-g)^{-1/2}[(-g)^{1/2}\Gmn\eta\der{\m}]\der{\n}=0,
 \label{jutreq}\eeqray
where $\Gmn$ is the inverse of the matrix
$\gmn=(\r/c)^{2/(D-1)}q_{\sss\mu \nu}$ with $q\_{00}=-(c^2-v^2),
~~q\_{0i}=q\_{i0}=- v\_{i},~~q\_{ij}=\d\_{ij}$, with $\r$ the
background flow density, $\vv$ its velocity, and $c$ the local
speed of sound ($g=-[\r^{(D+1)}/c^2]^{2/(D-1)}$ is the determinant
of $\gmn$). The matrix $\gmn$ is called the acoustic space-time
metric because eq.(\ref{jutreq}) is identical to the
massless-scalar wave equation in a curved space time described by
the metric $\gmn$, $\Box$ being the covariant d'Alembertian. This
setup is used to simulate the propagation of light in
gravitational fields. The analogy is, however, anything but
complete. For example, a coordinate transformation of an acoustic
metric does not take us to another acoustic metric. And, in the
flat-space-time analog (homogeneous background flow at rest) there
is no parallel with observer independence of the speed of light.
The situation is more akin to propagation of light in the old
aether. Also, there is not yet an analog of the Einstein equations
whereby the effective geometry is determined by its sources.
Still, the analogy, where it does exist, is very useful and
captures some crucial aspects of photon propagation in curved
space times. For example, it elucidates the behavior of light near
event horizons in ``black hole'' geometries. It can also model
Lorentz invariance breakdown in light propagation, etc. (see,
e.g., \cite{unruh95,jacobson91,visser98}).
\par
Here I propose to extend this analogy to massive particles, also
in the context of ideal fluids, in the usual hope that it might
teach us about the real processes it represents. The interest in
such models might be twofold. First, they provide models for
emergent relativistic inertia: Starting with objects that have
negligible inertia of their own, their interaction with the fluid
puts a cost on their motion by endowing them with an effective
kinetic action and thus with pseudo-energy and pseudo-momentum.
Such models might shed light on the origins of real inertia, in
the spirit of Mach's principle. (I mean here Mach's principle in
the extended sense that inertia is not an innate attribute of
bodies but emerges as a result of their interaction with some
omnipresent agent, such as a field, the vacuum, the fluid in our
case, or, as in the original view, the totality of other bodies in
the universe.) They also permit us to study possible mechanisms
for breakdown of the standard Lorentzian kinematics at high
Lorentz factors. We can also study possible departures from
standard inertia when the global setup of the fluid is changed to
mimic real inertia in the context of the non-trivial cosmology of
our universe. In fact, my own interest in the subject stemmed
originally from the wish to construct mechanical models for
modified inertia that will mimic the behavior of MOND, a theory
that I proposed to replace the need for dark matter in galactic
systems (e.g. \cite{milgrom83,milgrom94a,sm02}).

\par
Second, these models extend the usefulness of acoustic analogs of
light propagation in curved space times to that of massive
particles in gravitational fields. Interestingly, I find that for
properly defined particles the same attribute that plays the role
of inertial mass also plays the role of passive gravitational mass
thus conforming to the weak equivalence principle. With these
models we can study mechanisms for the breakdown of the weak
equivalence principle, dynamics of massive particles near black
hole analogs, such as the existence  of a last stable orbit, etc..
And, with an appropriate definition of the particle vacuum we may
be able to study Hawking radiation and other quantum effects in
curved space time for massive particles.

\par
The particles I shall describe do not generally follow geodesics
of the acoustic metric itself, for which the proper time is
 \beqray d\tau=\a\c^{-1}dt,  \label{huitda} \eeqray
where $\c=\{1-[d\vec x/dt-\vv(\vr)]^2\}^{-1/2}$ is the ``Lorentz
factor'' of the velocity relative to the fluid, and $\a\equiv [\r
c^{(D-2)}]^{1/(D-1)}$. In the real world Lorentz invariance
dictates the above path length as the particle action; but this is
not so in the fluid context. Nevertheless, it would still be
useful to find analogs that have enough of the properties of real
particles, in particular, relativistic, quasi-Lorentzian dynamics.
This I begin to do in this paper. As I shall show, there is, in
fact, a generalization of the acoustic metric in the form of a
Finslerian one whose path length is the particle action and which
shares the null geodesics with the acoustic metric; so, a unified
description of massive and massless particles does emerge with a
Finslerian acoustic metric.
\par
I am not concerned here with the practicability of laboratory
construction of such analogs. I view their usefulness mainly as
theoretical laboratories for testing ideas concerning inertia and
gravity.
\par
A rather different approach towards mimicking massive particles in
the context of  Bose-Einstein condensates is described in
\cite{vw04,vw05}. The acquisition of induced mass by vortices
moving in superfluids has been discussed in \cite{volovik03} and
references therein.
\par
 In section II I discuss the general idea
and define the particles. Section III contains the derivation of
the effective particle action in $D$ dimensions. In section IV I
discuss various aspects of the resulting dynamics of the
particles, first in flat space times, then in the presence of
analog gravity. Section V brings up some additional issues.

\section{Massive particles in flat space time}

\par
Very weak perturbations of the fluid flow itself, to wit acoustic
waves, are the analog of light in the fluid context. They are
described by the same degrees of freedom as the unperturbed,
background flow and move with the local speed of sound relative to
the fluid. Analogs of massive particles should be able to move at
any ``subluminal'' speed relative to the fluid. They should thus
be defined as regions of space where the equations of motion for
the background flow break down. The exact definition of the
particles, with the prerequisites they have to satisfy, is best
presented in the context of ``flat'' space times; i.e.,
homogeneous fluids at rest.
\par
A rigid body moving with constant speed in an inviscid,
incompressible fluid is subject to no force; this is known as the
d'Alembert paradox. It follows straightforwardly from scaling that
the energy of the fluid in this configuration is proportional to
the body's velocity squared. Accelerating the body increases the
energy of the fluid hence creating an effective force resisting
acceleration. The force is proportional to the acceleration,
giving rise to acquired Newtonian inertia: The speed of sound in
incompressible fluids is formally infinite, so there is no
velocity parameter to represent the speed of light. The effective
inertial mass tensor is the fluid density times some geometrical
volume matrix that depends on the shape of the body (See e.g.
\cite{lamb32} and \cite{ll87} \S 11). My aim here is to extend
this idea to bodies with relativistic kinematics. Naturally one
then begins with a compressible background fluid whose speed of
sound will play the role of the speed of light. A rigid body is
however not a good model for a particle (see below) so we'll have
to find others.
\par
A more general discussion of forces on static bodies in a class of
nonlinear media, of which our simple flow is an example, can be
found in \cite{milgrom02} where I consider different possible
definitions of bodies. For example, to define a body one can
dictate boundary conditions on a closed rigid surface of a region
that can move in the fluid. Dictating a vanishing normal component
of the flow velocity defines a rigid body, for instance.
Alternatively, we can define the particle as a rigid collection of
sources (and sinks). Yet another way is to dictate inside the
particle an external potential that couples to the flow density.
And yet another is to take the particle as a small region of
non-vanishing vorticity. There are more options; the choice,
however, is limited by the following requirements that I think
should apply in the quest of analog massive particles: a. The
particle should constitute a controllably weak perturbation on the
background flow. This is not just to facilitate the derivation of
the particle's dynamics but mainly to prevent the particle from
probing the equation of state of the fluid at densities other than
the background value. This requirement eliminates, for example,
rigid bodies as candidate particles because when such bodies move
with relativistic speeds they create strong perturbations in their
vicinity, no matter how small in size they are. b. An effective
principle of inertia should hold: when the particle is set in
motion in a homogeneous fluid it should retain a constant speed;
i.e. it should not be subject to forces by the fluid. The
d'Alembert paradox insures this for a rigid body in an
incompressible ideal fluid. It was shown in \cite{milgrom02} that
this is true also for compressible fluids, and, in such a case,
also when the body is defined as a region of dictated external
potential or as a distribution of sources, provided the integrated
source out-flux vanishes (this holds exactly, even when the body
is not a weak perturbation). c. To limit the scope of the
discussion I also require in this paper that the particle is a
rigid object, with no internal degrees of freedom. It may be
interesting to relax this assumption in various ways.
\par
I shall indeed concentrate here on particles defined as either a
distribution of sources or a dictated potential. The former is
epitomized by a source-sink dipole such as a small pipe within
which there is a pump sucking fluid at one end and ejecting it at
the other, or by any arrangement of dipoles such as a dipole
source layer, etc.. The second type may be realized archetypically
as a set of electric charges held together rigidly by a structure
that does not disturb the fluid mechanically (a rigid cage), and
moving in a weakly charged fluid with constant charge-to-mass
density ratio. It is best to take the total body charge as zero so
as to attain a confined potential. Ideally it would be good to add
an inert (static) background with the opposite charge to cancel
that of the unperturbed fluid, so that only density perturbations
carry net charge.
\section{The effective particle action}
One can get the equations of motion of an irrotational, inviscid,
barotropic fluid in $D$ space dimensions from the effective action
 \beqray S=-\int\dDx[\r\f\dt + \oot\r(\gf)^2+e+\r\theta+ \r\psi+\f
 s], \label{action}\eeqray
  where $d^{D+1}x$
stands for $dt\drD$, $\r$ is the fluid density, $\f$ is the
velocity potential: $\vv=\gf$, and $e(\r)$ is the intrinsic energy
per unit volume, which is a function of $\r$ for a barotropic
fluid. The action $S$ is based on that derived in \cite{schakel96}
to which I have added a source term with source density
$s(\vr,t)$, and potential terms. The potential fields
$\theta(\vr,t)$ and $\psi(\vr,t)$ couple to the fluid density.
They can be of the same type but I write them separately because
they have different roles: $\theta$ is completely dictated
externally, and partakes in establishing the unperturbed
background flow, while $\psi$ represents a particle and so
constitutes a weak perturbation confined to a very small, freely
moving region of space. (I kept here the sign of the action, which
is derived in \cite{schakel96} from the fluid action $\int \r
v^2/2-e$; so it is clear with which sign to add actions for
additional degrees of freedom. Kinetic energies are added with a
plus sign, while potential energies, such as $\r\psi$, appear with
a minus sign.)
\par
 Varying the action over $\f$ gives the continuity equation:
 \beqray\r\dt+\div(\r\gf)=s.\label{continuity} \eeqray
Varying over $\r$ gives the Bernoulli equation:
\beqray\f\dt+{1\over 2}(\gf)^2+h(\r)+\theta+\psi=0,
\label{bernoulli}\eeqray
 which in the barotropic irrotational case is equivalent to the
Euler equation. Here $h(\r)=e'(\r)$ is the specific enthalpy.
\par
 To the fluid degrees of
freedom we now add those of the model particle: its position
$\rs(t)$, and possibly its orientation. At this stage I want to
eliminate the orientation as an unnecessary (but possibly
interesting) complication. Later on I shall assume a spherically
symmetric particle for which this is not an issue. For the time
being I shall take an arbitrarily shaped particle but assume that
its orientation is kept fixed in space (e.g. by providing it with
a gyroscope), which generically gives anisotropic inertia. The
particle's dynamics will turn out to depend on its orientation
with respect to its velocity relative to the fluid. If we can
somehow keep this orientation fixed (e.g. by providing the body
with efficacious fins) dynamics will be isotropic for any body
shape.
\par
  One type of particle I treat is a
small region of space, positioned around $\rs(t)$, where a rigid
arrangement of sources is dictated. This corresponds to a source
distribution \beqray s(\vr,t)=\hs[\vr-\rs(t)]. \label{huia}
\eeqray $\hs$ vanishes everywhere except in a volume of diameter
$a$ much smaller than any length scale characterizing the
unperturbed flow and the trajectory of the particle, and
corresponds to a vanishing total outflow:
\beqray\int\hs(\vr)\drD=0.\label{nuy}\eeqray
 A different type of particle may be
represented by a small volume of diameter $a$ around $\rs(t)$
where an external potential is dictated:
\beqray\psi(\vr,t)=\hpsi[\vr-\rs(t)]\label{mio} \eeqray
 with $\hpsi$ vanishing rapidly beyond the radius of the source.
While $\hs(\vr)$ or $\hpsi(\vr)$ are fixed and constitute the
internal structure of the particle, its position $\rs$ is free. My
aim is to derive an effective action for $\rs$ by solving for $\r$
and $\f$ for a given trajectory $\rs(t)$, then substitute these
back in the action to get a functional of $\rs(t)$ that is the
required effective action of the particle. I do this under the
assumption that the particle is a weak perturbation on the
background flow and is very small.
\par
Let us keep the designation $\r$, $\f$ for the unperturbed,
background flow attributes, and write the density and velocity
potential in the presence of such perturbations as $\r+\z$ and
$\f+\eta$, respectively. Expanding the action to second order in
$\z$ and $\eta$ we get a zeroth order term, which is taken as a
constant for a given background. The first order term is, after
some integrations by parts,
 \beqray S_I&=&-\int
d^{D+1}x\{\z[\f\dt+\oot\gfs+h(\r)+\t]\nonumber\\&-&\eta[\r\dt+\div(\r\gf)]
+(\r\eta)\dt+\div(\eta\r\gf)\}\nonumber\\&+&\int \dDx (\r\psi+\f
s).     \label{yumba}\eeqray
 The first two terms vanish for
solutions of the unperturbed field equations. The next two terms
are the usual integrals of complete derivatives; they vanish if we
can neglect the perturbation at space and time infinities. The
second integral may engender first order effective forces on our
particle and I want to eliminate it. For a potential particle this
can be done by assuming a background flow of constant density, as
I shall eventually assume anyway. In this case this term becomes
an immaterial constant contribution to the Lagrangian $\propto
\r\int\drD\hpsi$. If we want to permit a variable density we add
to the background flow an inert background distribution of charges
that cancels that of $\r$, then $\psi$ couples only to the
perturbation $\z$ and this first order term disappears.
  For a source particle this first order term
is analogous to the energy of an electric charge distribution $s$
in a potential field $\f$. For a spherical particle, such as I
treat most of the time, this term actually vanishes for a constant
background density. In this case the continuity equation is the
Laplace equation for $\f$. Then write $\hs\propto \Delta\Phi$,
integrate twice by parts, and note that because $\hs$ is spherical
and of vanishing integral, Gauss's theorem says that $\Phi$
vanishes outside the source. (Our charge distribution produces no
field of its own outside it. The external field, which satisfies
the Laplace equation, has no sources inside the charge
distribution, so the source does not interact with the field.) If
the particle is not spherical, expand the space integral in this
term in multipoles about $\rs$. The monopole contribution
$\f(\rs)\int\hs(\vr)\drD$ vanishes. The dipole term
$\vv(\rs)\cdot\int\hs(\vr)\vr\drD$ is of the same order in the
particle size as the terms I shall want to keep (see below) and I
get rid of it by taking particles with vanishing dipole. The
quadrupole $v\_{i}\der{j}(\rs)\int \hs(\vr) r_i r_j\drD$, and
higher multipoles, is of higher order in the particle size (while
being of a lower order in the perturbation) and I neglect such
contribution to the action because after the strength of the
perturbation is set we can take the particle size as small as we
wish.

\par
Turn now to the second order action, from which the effective
particle action is constructed. We have

\beqray S_{II}=&-&\int
d^{D+1}x[\z\eta\dt+\oot\r(\grad\eta)^2\nonumber\\
&+&\z\gf\cdot\grad\eta+{c^2\over 2\r}\z^2+\z\psi+\eta s],
\label{actionII}\eeqray with $c$ the speed of sound: $c^2\equiv
p'(\r)=\r h'(\r)$).
 It gives the first order Bernoulli equation
 by varying over $\z$:
 \beqray\eta\dt+(c^2/\r)\z+\gf\cdot\grad\eta+\psi=0, \label{berI}\eeqray
and the continuity equation,
 \beqray\z\dt+\div(\r\grad\eta)+\div(\z\gf)=s, \label{contI}\eeqray
by varying over $\eta$.  The first can be used to eliminate
 \beqray \z=-(\r/c^2)(\eta\dt+\gf\cdot\grad\eta+\psi),\label{iuop}\eeqray
and substituting in the second we get
  \beqray
&-&[(\r/c^2)(\eta\dt+\gf\cdot\grad\eta+\psi)]\dt+\div(\r\grad\eta)
\nonumber\\&-&\div[(\r/c^2)(\eta\dt+\gf\cdot\grad\eta+\psi)\gf]=s.
\label{i} \eeqray

Rearranging gives
 \beqray
&-&[(\r/c^2)(\eta\dt+\gf\cdot\grad\eta)]\dt+\div(\r\grad\eta)
\nonumber\\&-&\div[(\r/c^2)(\eta\dt+\gf\cdot\grad\eta)\gf]
\nonumber\\&=&s+[(\r/c^2)\psi]\dt+\div[(\r/c^2)\psi\gf].
\label{ii} \eeqray
 I use the continuity equation for the background flow to write
$s+\r(\psi/c^2)\dt+\r\gf\cdot\grad(\psi/c^2)$ for the right hand
side. The left hand side is $\gh\Box\eta$, where, as in
eq.(\ref{jutreq}), $\Box$ is the covariant d'Alembertian
corresponding to the acoustic metric $\gmn$. So

\beqray \Box\eta=\tilde
s+(\psi/c^2)\der{\m}J^{\m},\label{eta}\eeqray
 where $\tilde s\equiv \gmh s$ is the covariant source
density, and the current
 \beqray J^{\m}\equiv\gmh\r(1,\vv)\label{current} \eeqray
 is covariantly conserved:
 \beqray J^{\m}\cd{\m}=\gmh[\gh J^{\m}]\der{\m}=0.
 \label{oplita}\eeqray

Now eliminate $\z$ from the action $S_{II}$ itself. After some
algebra, one gets inside the integral (up to a total derivative)
\beqray S_{II}&=&-\int \dDx\gh[{1\over
2}\eta\der{\m}\eta\der{\n}\Gmn+\eta
(\psi/c^2)\der{\m}J^{\m}\nonumber\\&+&\eta\tilde s-{1\over 2\r
c}\psi^2].\label{siifin} \eeqray
 Varying over $\eta$ gives the field equation (\ref{eta}).

\par
The program is then as follows: for a given  $\rs(t)$, which,
together with the given $\hs$ or $\hpsi$, determines the source
term, solve eq.(\ref{eta}) for $\eta(\vr,t)$, then substitute it
in the expression for $S_{II}$ to get the value of the effective
action as a functional of the trajectory; this is the particle
action we are after, $S[\rs(t)]$. Equation.(\ref{siifin}) requires
knowledge of the solution $\eta$ everywhere in space time. A more
manageable expression is gotten by employing the integral relation
 \beqray \int \dDx\gh[\eta\der{\m}\eta\der{\n}\Gmn+\eta
(\psi/c^2)\der{\m}J^{\m}&+&\eta\tilde
s]\nonumber\\&=&0,\label{sukas} \eeqray
 which holds for solutions of the field equation up to surface terms at
space-time infinity. (This is a simple special case of the results
of \cite{milgrom94b} and follows straightforwardly by integrating
the first term by parts to give $-(1/2)\eta\Box\eta$ then using
the field equation (\ref{eta}).) So we can set
 \beqray
S[\rs(t)]=&-&\int \dDx\gh[\oot\eta
(\psi/c^2)\der{\m}J^{\m}\nonumber\\&+&\oot\eta\tilde s-{1\over 2\r
c}\psi^2].\label{sipeq} \eeqray
 This expression requires knowledge of $\eta$ only inside
the particle, where either $\psi$ or $\tilde s$ don't vanish; this
is very helpful.
\par
It is impracticable to solve for $\eta$ for an arbitrary
trajectory in an arbitrary background flow. It is clear that the
resulting effective action would be time non local.
 However, the assumed smallness of the particle permits us to approximate $\eta$ inside
the particle in a way that depends only on the instantaneous state
of motion, and this will result in a local approximation of the
action. For a very small particle we can assume that as it moves
about, a steady state corresponding to the instantaneous
conditions is reestablished within the particle on the short time
scale it takes sound waves to get from one end of it to the other.
We essentially separate the dependence on macroscopic coordinates
and the microscopic ones within the body, where $\eta$ changes
quickly, by assuming that from eqs.(\ref{huia}) and (\ref{mio}) we
can write to a very good approximation (becoming exact in the
limit of infinitesimal particle size)
 \beqray\eta(\vr,t)=\heta[\vr-\rs(t)] \label{miou}\eeqray
to describe the fast variations of $\eta$ around the particle's
position in space time, and where $\heta$ still depends on
macroscopic properties such as the flow and particle velocities
and the fluid density at $\rs(t)$. I shall discuss below the
conditions for this approximation to hold.
\par
We now calculate $\Box\eta$ with this ansatz. I again make use of
the fact that, due to the smallness of the particle, the space and
time variations of $\eta$ are dominated by those produced by the
fast variations in $\heta[\vr-\rs(t)])$. So, for example, in
$\eta\dt=-\vvs\cdot\grad\heta+q$, where $q$ represents terms
coming from the implicit dependence of $\heta$ on macroscopic
quantities and their time variation, we neglect all such terms.
Then in $\eta\dt\dt\approx (\vvs\cdot\grad)^2\heta -(d\vvs/dt)
\cdot\grad\heta$ (again neglecting $q$ terms)
 I further neglect the second term (by our approximation
$|d\vvs/dt|\ll v_*^2/a$ generically, $a$ being the diameter of the
particle). With this approximation, which leaves us only with
terms with second derivatives of $\heta$, we get

\beqray \Box\eta={c\over \r}\{\Delta-[{\vU\over c}
\cdot\grad]^2\}\heta, \label{jlse} \eeqray

where, $\vU\equiv\vvs-\vv$ is the relative velocity of the
particle with respect to the fluid, and $\vv, ~\r, ~c$ are
evaluated at $\rs(t)$. Thus
 \beqray \heta\der{x}\der{x}+\heta\der{y}\der{y}+\c^{-2}\heta\der{z}\der{z}=
 \r^{-1}\hs-(U/c^2)\hpsi\der{z}, \label{mseui}\eeqray
  where $\c$ is the relative ``Lorentz factor''
 $\c=(1-U^2/c^2)^{-1/2}$, the $z$
axis is in the direction of $\vU$, and where I used the fact that
now $J^\m(\psi/c^2)\der{\m}
=\gmh(\r/c^2)(\psi\der{t}+\vv\grad\psi)=-(\r
c)^{-1}\vU\cdot\grad\hat\psi $.
 In the coordinates $x'=x, ~y'=y, ~z'=\c z$ eq.(\ref{mseui})
 becomes the Poisson equation
 \beqray \Delta'\heta=\r^{-1}\hs[\vR(\vec R')]-Uc^{-2} \hpsi\der{z}[\vR(\vec R')],
 \label{poiss}\eeqray
 provided the relative speed $\vU$ is subsonic.
This appearance of the stretched Laplacian in the linearized
equation for weak perturbations moving with subsonic speed in a
compressible fluid is familiar, for example, from the treatment of
a constant flow past a thin wing very nearly parallel to the flow
(e.g. \cite{ll87} \S 124). The perturbation there enters not
through source terms as here, but through the boundary conditions
on the rigid wing, leaving us with a distorted Laplace equation
instead of Poisson's as here. This equation is elliptical for
subsonic speeds for which our treatment below applies, but become
hyperbolic for supersonic speeds.

\par
The effective particle action can then be written as

\beqray S[\vr_*(t)]=\int L~dt, \label{kodt} \eeqray
 with the particle Lagrangian
 \beqray L=&-&\oot\int\drD~\heta(\vr)\hs(\vr)+\oot{\r \over c^2}
 \int\drD~\heta(\vU\cdot\grad)\hpsi\nonumber\\&+&\oot{\r\over c^2}\int\drD~\hpsi^2,
  \label{pointo} \eeqray
where (for $D>2$)
  \beqray \heta(\vr')=-{1\over (D-2)\OD}\int \dRD'{\r^{-1}\hs
 -c^{-2}(\vU\cdot\grad)\hpsi\over |\vr'-\vec
 R'|^{D-2}} \label{greens} \eeqray
 is the solution of the stretched Poisson equation;
  $\OD$ is the solid angle in $D$ dimensions. The $D=1,2$ cases
will be treated separately in Appendix A.
\par
 To recapitulate, the approximation I made amounts to the
 following procedure:
At any given time take the local values of the velocities of the
fluid and the particle and of the fluid density, calculate the
steady state solution, $\heta$, from eq.(\ref{mseui}) for a
homogeneous fluid with these properties and an eternally constant
particle velocity, then use this for the instantaneous $\eta$
inside the particle.
\par
I now proceed to discuss separately source and potential
particles.
\subsection{Source particles}
 For a pure source particle put $\psi\equiv 0$; then
substituting expression (\ref{greens}) in eq.(\ref{pointo}) and
changing to the $\vr'$ variables we get (for $D>2$)

 \beqray L= {1\over 2(D-2)\OD\r\c}\int \drD'
\dRD'{\hs[\vec R(\vec R')]\hs[\vr(\vr')]\over |\vr'-\vec
 R'|^{D-2}}.\label{kp} \eeqray

The integral in eq.(\ref{kp}) is proportional to the
``electrostatic'' energy of a charge distribution $\hs$ stretched
by a factor $\c$ in the $z$ direction. Note that $L$ is positive
because it is proportional to
$-\int\heta\Delta\heta=\int(\grad\heta)^2$ ($-\Delta$ is a
positive definite operator).

We can also write the integral in terms of the $\vr$ coordinates
as \beqray &L&= {\c\over 2(D-2)\OD\r}\times\nonumber\\&&\int {\drD
\dRD~\hs(\vec R)\hs(\vr)\over \{(\vr-\vec
R)^2+\c^2[(\vU/c)\cdot(\vr-\vec R)]^2\}^{(D-2)/2}} ,\label{kpiop}
\eeqray
 where the full dependence of $L$ on $\c$, the structure of
the particle, and its orientation with respect to the relative
velocity is explicit.
\par
In the non-relativistic limit, $U\ll c$, eq.(\ref{kpiop}) tells us
that

\beqray L={E_0\over \OD\r}+\oot U_im_{ij}U_j+O(U^4/c^4);
\label{poyta} \eeqray

$E_0=[2(D-2)]^{-1}\int \drD \dRD~\hs(\vec R )\hs(\vr)/ |\vr-\vec
R|^{(D-2)}$ is the value for the unstretched configuration, and
the effective mass tensor is

\beqray m_{ij}&=&{1\over 2(D-2)\OD\r c^2}\times\nonumber\\&&\int
\drD \dRD{\hs(\vec R )\hs(\vr)\over |\vr-\vec
R|^{D-2}}\times\nonumber\\&&[\d_{ij}-(D-2){(r-R)_i(r-R)_j\over
(\vr-\vec R)^2}].\label{kpopa} \eeqray

In the isotropic case (for which a cubic symmetry of the particle
suffices) we get the mass of the particle

\beqray m=Tr(m_{ij})/D={2E_0\over D\OD\r c^2}.  \label{jukaga}
\eeqray

When the background density is not a constant of the configuration
this mass parameter is a function of space-time position through
$\r$ and possibly $c$. I shall still refer to it as the mass of
the particle.

To insure isotropy of inertia I shall assume henceforth that our
particle is spherically symmetric. In this case $L$ can be
obtained analytically. This is done in Appendix A and yields
 \beqray L=L_0F(1,\oot;{D\over 2};{U^2\over c^2}), \label{byutda} \eeqray
where $L_0=E_0/\OD\r$ is the value of the effective Lagrangian for
$U=0$, and $F$ is the Gauss hypergeometric function.

\subsection{Potential particles}

 Consider now a pure
potential particle ($\hs=0$). Repeating the same argumentation as
before

\beqray L&=& {\r \over 2(D-2)\OD
c^4\c}\times\nonumber\\&&\int\drD' \dRD'
{\vU\cdot\grad_R\hpsi[\vec R(\vec
R')]\vU\cdot\grad_r\hpsi[\vr(\vr')]\over |\vr'-\vec
R'|^{D-2}}\nonumber\\&+&\oot{\r\over
c^2}\int\hpsi^2\drD.\label{kk} \eeqray
 Or, with derivatives with respect to $\vr'$ and $\vR'$,

\beqray L&=& {\r \c\over 2(D-2)\OD
c^4}\times\nonumber\\&&\int\drD' \dRD'
{\vU\cdot\grad_{R'}\hpsi[\vec R(\vec
R')]\vU\cdot\grad_{r'}\hpsi[\vr(\vr')]\over |\vr'-\vec
R'|^{D-2}}\nonumber\\&+&\oot{\r\over
c^2}\int\hpsi^2\drD,\label{kh} \eeqray

In the non-relativistic limit

\beqray L=\oot{\r\over c^2}\int\hpsi^2\drD+\oot
U_im_{ij}U_j+O(U^4/c^4), \label{koiuts} \eeqray

where the mass tensor is (integrating by parts)

\beqray m_{ij}&=&{\r\over (D-2)\OD c^4}\times\nonumber\\
\int&\drD& \dRD \hpsi(\vec R)\hpsi(\vr)
\pd{}{R_i}\pd{}{r_j}|\vr-\vec R|^{-(D-2)}.\label{kkasta} \eeqray

In the isotropic case

\beqray m=Tr(m_{ij})/D={\r\over
Dc^4}\int\hpsi^2\drD,\label{kkahuy} \eeqray
 where I used $-\Delta[r^{-(D-2)}]=(D-2)\OD\d^D(\vr)$.
 The first integral
in eq.(\ref{kh}) may be viewed as proportional to the energy of a
polarized medium with unidirectional polarization of magnitude
$\propto U\hpsi[\vr(\vr')]$. It can be calculated exactly for an
arbitrary, spherically symmetric distribution $\hpsi(r)$. The
integral is calculated in Appendix B, and when added to the second
integral we get
 \beqray L&=&{\r \int\hpsi^2 \drD\over 2c^2}F(1,\oot;{D\over 2};{U^2\over c^2})
 \nonumber\\&=&{D\over 2}mc^2F(1,\oot;{D\over 2};{U^2\over c^2}),\label{mushta}\eeqray
 with the same hypergeometric function appearing in the
Lagrangian of a source particle. These identical results are
obtained from rather different starting expressions, and I have
not been able to find an underlying physical reason for the
equality.

\subsection{Aspherical and compound particles}
A larger variety of $\c$ dependences of $L$ is afforded by
considering aspherical particles. If the orientation of the
particle is kept fixed in space, anisotropic inertia results
generally; but, if we can somehow keep the orientation fixed with
respect to $\vU$, the effective inertia is isotropic.
\par
As an example, consider a hyper-planar, bipolar source layer
(charged planar capacitor in the electrostatic analog) whose
normal always makes an angle $\Theta$ with the relative velocity
vector. (Actually, because of the reservations discussed above, we
need to take two, back-to-back dipole layers to annihilate the
dipole moment of the particle, but this is immaterial for the
results since the two layers do not interact, so I shall just
continue to speak of one layer.) From eq.(\ref{kp}) the Lagrangian
is $\c^{-1}E_c$, $E_c$ being the energy of the stretched bilayer.
This energy, like that of a charged capacitor, is $E_c\propto Q^2
d/A$, where $Q$ is the total charge on one layer, $A$ the area,
and $d$ the spacing. Under stretching $Q\rar \c Q$, $A\rar
A(cos^2\Theta+\c^2 sin^2\Theta)^{1/2}$, and $d\rar d\c
(cos^2\Theta+\c^2 sin^2\Theta)^{-1/2}$. So,
 \beqray L=L_0{\c^2\over
cos^2\Theta+\c^2 sin^2\Theta}={L_0\over
1-(U^2/c^2)cos^2\Theta}.\label{nutata}\eeqray
 This varies between $L=L_0 \c^2$ for $\Theta=0$, as in the 1-D
case, to a constant $L=L_0$ when the bilayer moves parallel to
itself relative to the fluid. Integrating over angles with weight
$sin^{D-2}\Theta d\Theta$ gives back our result for the spherical
case.  (In the relativistic limit the contribution to the action
of a spherical particle then comes from a small $\Theta$ region
near the leading point on the sphere. The corresponding area
decreases with increasing dimension; hence the strong $D$
dependence of the relativistic limit.) All the above applies to
any collections of bi-layers making the same angle with the
relative velocity vector, for example a cone of half-opening angle
$\pi/2-\Theta$ moving always along its axis relative to the fluid,
like an arrowhead. In general, if we tie together several
particles that are so far from each other that their mutual
interactions can be neglected compared to their self interactions,
the body will have an effective Lagrangian that is the sum of
those of the individual components. (The inter-component distance
still has to be small compared with the scale over which
macroscopic properties of the flow vary).

\subsection{Limitations and caveats}
I made two types of assumptions about the particle: that it
constitute a very weak perturbation on the fluid even within the
particle itself, and that it is very small in size so that
conditions in it very quickly take up steady state values
corresponding to the momentary macroscopic conditions around it. I
now discuss in more detail what these require from the strength of
the potential $\hpsi$, from the source density $\hs$, and from the
particle size $a$. I find that in general these approximations
break down at high $\c$ however small $a$ and $\hpsi$ or $\hs$
are. The perturbation treatment assumes that $\z\ll\r$.
Equation(\ref{iuop}) gives
 \beqray
 \z&=&-(\r/c^2)(\eta\der{t}+\vv\cdot\grad\eta+\psi)\nonumber\\&\approx&
 (\r/c^2)(\vU\cdot\grad\heta-\hpsi)
 ,\label{guytrew}\eeqray
 where I used our approximation in the second equality. Consider
 first a potential particle.
For $\vU\rar 0$, $\eta\rar 0$ and the basic requirement is
$|\z|/\r\approx |\psi|/c^2\ll 1$. We further have to insure that
for the limit $U\rar c$ we still have $\z/\r\ll 1$, so we need
  $|\vU\cdot\grad\heta|/c^2\ll 1$ everywhere in the body.
The exact constraint this puts on $\c$ depends on the particle
structure. We can get an estimate of this quantity by noting that
what I calculated as the first term contributing to $L$ is
$(\r/2c^2)\int\drD\heta(\vU\cdot\grad)\hpsi=-(\r/2c^2)\int\drD\hpsi(\vU\cdot\grad)\heta$.
If we take, for example, a particle of constant $\hpsi$ we know
the value of $(\vU\cdot\grad)\heta$ is constant inside the
particle and from the results for $L$ it is
 \beqray \vU\cdot\grad\heta=-\hpsi[F(1,\oot;{D\over 2};{U^2\over
c^2})-1].\label{juipota} \eeqray
 So we also need
 \beqray (|\hpsi|/c^2)[F(1,\oot;{D\over 2};{U^2\over
c^2})-1]\ll 1.\label{juipura} \eeqray
 In the non-relativistic limit the expression in square
 parentheses behaves as $U^2/c^2$, so no new requirement is added.
 In the relativistic regime we
can get a validity limit on $\c$. For $D>3$ the $F$ above is
finite for $U=c$
 and we do not get an additional constraint; the basic one suffices
 for all values of $\c$.
For $D=1$ we have to have $\c^2|\hpsi|/c^2\ll 1$, for $D=2$:
$\c|\hpsi|/c^2\ll 1$, and for $D=3$: $ln(\c)|\hpsi|/c^2\ll 1$. For
a source particle we have, as before, to first order in the source
strength $\z\approx
 (\r/c^2)\vU\cdot\grad\heta$, which vanishes when the particle is
at rest with respect to the fluid. Consider such a particle in a
static fluid. Neglecting the variation of the external potential
across the particle, the Bernoulli equation tells us that $\z$ is
indeed second order in the source. If we write $\hs=\r\Delta\Phi$
we have from the continuity equation that to first order
$\heta=\Phi$ (with 3rd order corrections), and so
$\zeta\approx-(\r/2c^2)(\grad\Phi)^2$.
 The basic requirement from $\hs$ for our approximation to hold
is then $|\grad\Phi|\ll c$. And here too, for $\c\gg 1$  we have
to have $|\vU\cdot\grad\heta|/c^2\ll 1$, which casts a constraint
on $\c$ that may depend on the particle structure and the
dimension.

\par
Consider now the condition on the particle size. When the
particle, having diameter $a$, is moving with velocity $\vU$
relative to the fluid, it takes a sound wave time
 $\d t\approx (a/c)(1-U/c)^{-1}=(a/c)(1+U/c)\c^2$ to move from the
aft of the particle to its fore. We want $\d t$ to be much shorter
than any time scale, $T$, over which the environmental parameters
change. The basic requirement, which should hold even at low
relative velocities is then $a\ll cT$.  In the relativistic regime
the requirement is that $a\ll cT/2\c^2$.
\par
We thus see that even for very small values of $\hpsi$ or $\hs$,
and of $a$, our approximations, and thus our results, are not
valid for $\c\rar \infty$. This is to be expected: For $D\le 5$
the particle energy diverges for $\c\rar \infty$, which, if valid
indefinitely, says that we cannot accelerate our particles to
supersonic speeds; but this is clearly not true.
\par
It would be interesting to see how our acquired dynamical
properties of particles are modified when the approximations break
down, as some of these may also be taking place in reality. For
example, the time locality of the Lagrangian is only a result of
the approximation: The effects of the particle on the fluid at one
time affect, at some level, the motion of the particle at another
time. Indeed they may affect other particles as well, thus
creating an effective interaction between particles mediated by
the fluid akin to the Cooper pairing interaction between electrons
in a superconductor. We also have here some ready made mechanisms
for the breakdown of Lorentzian dynamics at high $\c$. All these
departures are still considered in the context of inviscid,
irrotational, barotropic fluids; and these attributes are also
only approximations (see a discussion of these in the context of
phonon propagation in \cite{visser98}).

\par
I have also neglected the goings on inside the source itself. This
is after all some parallel flow (e.g. in some pipes with pumps)
that move the fluid from the sinks to the sources.

\section{Particle dynamics}
Notwithstanding the absence of true inertia in our particles, they
 acquire relativistic inertia through their interaction with the
fluid; this is encapsuled in the kinetic action for free
particles.
\par
From now on I shall assume a position and time independent density
for the background flow.  This situation is rather less cumbersome
to describe, captures most of the concepts I want to introduce,
and insures a constant speed of sound, which after all is our
analog of the speed of light (this can also be insured, without
imposing a constant $\r$, by having a fluid equation of state of
the form $p=c^2\r+const.$). It also means that $m$ is a constant
and so both types of particles have the same motion. (If the fluid
density depends on position or on time we get a variable mass for
the particles, which might be interesting to explore.) The freedom
left in selecting a background flow--a solution of the field
equations with $s=0$ and $\psi=0$--is then only in choosing the
velocity potential field from among the harmonic functions. We
then have to impose an external potential $\theta$ that will
satisfy the Bernoulli equation for the chosen velocity field. The
exact form of the equation of state is immaterial since we shall
probe it only at one density value where its derivative only
(speed of sound) has to be known.
\par
 For either definition of a particle the effective
action can be written in the form
 \beqray S=\int mc^2\ell[\c(t)]dt. \label{miouy}\eeqray
 For a source particle $m$ is
proportional to the ``electrostatic'' energy of a charged medium
with charge density $\hs$, and for a potential particle it is
proportional to the ``magnetostatic'' energy of a sphere with
unidirectional polarization $\hpsi$.
\par
I now proceed to discuss various aspects of the resulting
dynamics.
\subsection{ Flat-space-time dynamics: emergent inertia}

In a flat space-time; i.e., in a homogeneous background flow at
rest, the effective Lagrangian for the two types of particles
discussed here is
 \beqray L=mc^2(D/2)F(1,\oot;{D\over 2};{v_*^2\over c^2}), \label{jiuot} \eeqray
where $v_*$ is the particle velocity and $m$ is constant. Note
that $D$ here is determined by the symmetry of the particle, and
is not necessarily the dimension of the space in which it moves.
For example, a plane symmetric particle that moves only along its
normal has $D=1$. More generally, a particle in $N$ dimensions of
cylindrical symmetry having the symmetry of $S^D\otimes R^{(N-D)}$
whose velocity is in the $S^D$ subspace  corresponds to dimension
$D$.
\par
Using the formula for the derivative of the hypergeometric
function we get for the momentum
 \beqray \vec p=\pd{L}{\vvs}=m\vvs F(2,{3\over
2};{D+2\over 2};{v_*^2\over c^2}).\label{muopia} \eeqray
 The kinetic energy
 \beqray E_k&=&-L+L_0+\vvs\cdot\vec p=mc^2\{{D\over 2}[1-
F(1,\oot;{D\over 2};{v_*^2\over c^2})]\nonumber\\&+&{v_*^2\over
c^2}F(2,{3\over 2};{D+2\over 2};{v_*^2\over c^2})\},\label{muoqa}
\eeqray where I added a constant so as to make $E_k$ vanish for
$\vvs=0$.
\par
These energy and momentum are conserved--as follows from
N\"{o}ther theorem's related to the assumed time independence and
homogeneity of the fluid: If the particle is subject to a
conservative force
 derived from a potential $\xi$, we have to add $\int-\xi[\rs(t)] dt $ to
 the action we started with; so; the particle now satisfies
 $\deriv{\vec p}{t}=-\grad\xi,$ and $E_k+\xi$ is conserved.
And if we have some inter-particle forces $\sum \vec p_i$ is
conserved, and if these forces are derived from a potential again
the total energy is conserved.
\par
Of course, $E$ and $\vec p$ are not the real energy and momentum
of the particles; these were assumed to have no inertia of their
own so they can carry no energy and momentum. It is the stirring
of the fluid by the motion of the particles that puts a real
inertial cost to their motion. The rates of change of $E_k$ and
$\vec p$ equal the rates of change of the energy and momentum of
the fluid induced when the particle changes its velocity; they are
thus equal to the external power input and the external force
imparted to the particle. Such quantities are called pseudo-energy
and pseudo-momentum. These are often useful in the description of
motion of objects in homogeneous media with which they interact
(See \cite{peierls91} 2.4 and the review by \cite{stone00}). A
direct calculation of the energy and momentum of the fluid might
have also served, but it is impractical. Attempting to calculate
them even for the simple, steady-state configuration, with the
particle ever at constant velocity, gave me ambiguous results:
When these quantities are written as integrals of fluid attributes
over a volume that has to be taken to infinity, the results depend
on the shape of the integration volume. This is similar to what
Peierls \cite{peierls91} finds when trying to calculate these
quantities for a sound wave. The present approach of proceeding
through the action seems to be the proper way to proceed.
\par
Since  $F(1,\oot;{D\over 2};{v_*^2\over
c^2})=1+D^{-1}v_*^2/c^2+O(v_*^4/c^4)$, we have in all dimensions
the non-relativistic behavior
 \beqray L-L_0\approx \oot mv_*^2,~~~\vec p\approx m\vvs, ~~~E_k\approx
 \oot m v_*^2. \label{guptat} \eeqray
\par
In the highly relativistic regime the behavior depends strongly on
the dimension. Dimension $D=3$ is critical in some sense: Because
$F(a,b;c;z)$ is finite for $z=1$ when $c>a+b$,  $L$ is finite as
$\c\rar\infty$ for $D>3$, diverges logarithmically for $D=3$, and
diverges as a power of $\c$ for $D<3$. Dimension $D=5$ is another
critical dimension above which the energy and momentum remain
finite for $\c\rar \infty$; for $D=5$ itself these quantities
behave as $ln(\c)$ in this limit (see below), while for $D<5$ they
diverge as a power of $\c$.
\par
Following are the Lagrangian, the momentum, and the energy for
dimensions $D\le 5$ in closed forms with their relativistic
limits:

 For $D=1$
 \beqray L&=&mc^2\c^2/2,~~~~ \vec p=m\c^4\vvs,
 \nonumber\\&E_k&=mv_*^2\c^2(\c^2-1/2) \rar mc^2\c^4. \label{kiopata}\eeqray
 For $D=2$
 \beqray L&=&mc^2\c, ~~~~ \vec p=m\c^3\vvs,
 \nonumber\\&E_k&=mv_*^2{\c^2(\c^2+\c-1)\over \c+1}\rar mc^2\c^3. \label{gutiopa}\eeqray
 For $D=3$
  \beqray L=mc^2{3\over 4(v_*/c)}ln\left({1+v_*/c \over 1- v_*/c}\right). \label{fut} \eeqray

\beqray \vec p&=&m\vvs {3\over 2(v_*/c)^2}[\c^2-{1\over
2(v_*/c)}ln\left({1+v_*/c\over 1-v_*/c}\right)]\nonumber\\&\rar&
{3\over 2}m\vvs \c^2, \label{trgfd} \eeqray
 and
 \beqray E_k&=&{3\over 2}mc^2[1-{1\over (v_*/c)}ln({1+v_*/c\over
1-v_*/c})+\c^2]\nonumber\\&\rar&{3\over 2}mc^2\c^2. \label{lalala}
\eeqray
\par
The case  $D=4$ is particularly interesting. We can then write,
using formula 9.131.2 in \cite{gr45} (henceforth GR)
 \beqray &L&= 2mc^2F(1,\oot;2;{v_*^2\over
c^2})\nonumber\\&=&4mc^2[F(1,\oot;\oot;\c^{-2})-\c^{-1}F(1,{3\over
2};{3\over 2};\c^{-2})]. \label{utimba}\eeqray

 However, we have generally $F(1,b;b;z)=(1-z)^{-1}$, which gives
\beqray L&=&{4mc^2\over 1+\c^{-1}},~~~~~\vec p={4m\c\vvs\over
(1+\c^{-1})^2}, \nonumber\\ &E_k&=4mc^2\left(\c{\c-2\over
\c+1}+\oot\right). \label{utiusa}\eeqray
 The kinematics is quasi-Lorentzian and becomes Lorentzian in the
limit of high $\c$, with $L\approx Mc^2(1-\c^{-1}),$ $\vec
p\approx M\c\vvs$, and $E_k\approx Mc^2\c$, where $M=4m$.
\par
 For $D=5$, using formula 9.137.14 and then 9.121.1 in GR we can write
 \beqray &L&=mc^2(5/2)F(1,\oot;{5\over 2};{v_*^2\over c^2})
 \nonumber\\&=&{15\over 4}mc^2(v_*/c)^{-2}[1-\c^{-2}F(1,\oot;{3\over 2};{v_*^2\over
c^2})],\label{nutagi}\eeqray where
 \beqray F(1,\oot;{3\over
2};{v_*^2\over c^2})= {1\over 2(v_*/c)}ln\left({1+v_*/c \over 1-
v_*/c}\right)\label{miztapa}\eeqray
 is, in fact, the $D=3$ Lagrangian.
 So, for $\c\rar\infty$, $\vec p\approx (15/2)m\vvs ln(\c)$, $E_k\approx (15/2)mc^2ln(\c)$.

\subsection{Antiparticles and the particle vacuum}
If we wish to push the analogy beyond the dynamics of isolated,
ever-existing particles,  and discuss pair creation and
annihilation we need to define new notions. We have to identify
antiparticles, and we need to have a proper definition of the rest
mass of our particles. Also, unlike phonons, our particles are not
an organic part of the fluid; they cannot be created out of it if
not put in by hand. So, pairs will not spring out of the fluid
even under energetically favorable conditions, such as near event
horizons or in strong ``electric'' fields unless we prepare a
``vacuum'' that has the potentiality to beget them.
 \par
 For a particle given by some $\hpsi$, or
some $\hs$, it is natural to define the antiparticle as that given
by $-\hpsi$ or $-\hs$ respectively. They have the same $m$ as the
particles. If we take a particle and its antiparticle and
superimpose them on each other we get an object that does not
interact with the fluid, and in whose presence the energy is that
of a fluid devoid of particles. The particles themselves are still
there as they are some imposed external structures that do not
physically annihilate each other, they only cancel each others
influence on the fluid when they coincide. When one separates a
superimposed pair it increases the energy of the fluid by an
amount that we should identify as $2m_0c^2$, with $m_0$ the rest
mass of a particle. Let the two coalesce with no outside help, and
the fluid goes back to its initial state plus waves carrying the
released energy $2m_0c^2$ emitted as ``annihilation radiation''.
\par
I thus envisage the particle vacuum as a fluid filled with
superimposed (annihilated) pairs. In this state they do not affect
the fluid, and their introduction into the fluid does not cost in
energy. The partners of each pair stick together, up to
fluctuations, because it is energetically favorable. And now, if
the proper circumstances arise pairs can be created from this
vacuum and annihilated into it emitting phonons.
\par
What is then the rest mass of the particles? I have not been
concerned so far with the energy it costs to insert the particles
into the fluid but only with changes induced by their movement;
so, the rest mass cannot be read off the Lagrangian. To determine
the rest mass I considered in more detail a source particle whose
rest mass we should take to be $E_r/c^2$, where $E_r$ is the
energy required to introduce the source into the fluid at rest.
Compare the energy of the steady-state fluid configuration with
the source inserted with that of the configuration with an
homogeneous fluid at rest having the same total mass. I argue in
Appendix C that for a spherical source of vanishing total out-flux
$E_r=E_0/\r\OD$ and hence the rest mass is given by
 \beqray m_0={D\over 2}m. \label{juytara}\eeqray
Again, no new mass parameter is introduced and $m$ also determines
the rest mass.
\par
We can thus write the complete expression for the energy of a
source particle
 \beqray E &=& mc^2[D-{D\over 2} F(1,\oot;{D\over 2};{v_*^2\over
c^2})\nonumber\\&+&{v_*^2\over c^2}F(2,{3\over 2};{D+2\over
2};{v_*^2\over c^2})].\label{suftara} \eeqray

\subsection{ Curved-space-time dynamics}
When the acoustic space-time is not flat--i.e., when the
background flow velocity is not constant--all our particles fall
in the same way in a gravitational field thus obeying the weak
equivalence principle. This is non-trivial and might have well
been otherwise. For example, the first order terms that we
arranged to be absent could destroy universal free fall. But
barring such departures, which we saw can be avoided by properly
defining the setup, $m$ can be thought of as both the inertial
mass and the passive gravitational mass of the particle. (When the
background density is not constant the two types of particles see
two different, but conformally related, space times.)

\par
Our Lagrangian is of the form $L=L(U^2/c^2)$, which is also true
of the standard acoustic line element. These give the
Euler-Lagrange equation

\beqray \deriv{(L'U_i)}{t} +L'U_kv_k\der{i}=0. \label{cyui}
\eeqray
 Using the fact that $\vv$ is irrotational we have
 \beqray c^{-2}{L''\over L'}\vU\deriv{(U^2)}{t} +\deriv{\vvs}{t}-\oot\grad
 v^2=0.
\label{cyuimat} \eeqray
 Using the formula for the derivative of the
hypergeometric function we have for our actions
 \beqray L'=\oot mc^2 F(2,{3\over 2};{D+2\over 2};{U^2\over c^2}),\label{cyuiabu} \eeqray and

\beqray L''={3\over D+2}mc^2 F(3,{5\over 2};{D+4\over 2};{U^2\over
c^2}). \label{cyuspa} \eeqray
\par
In the limit $U\ll c$ the first term in eq.(\ref{cyuimat}) is of a
higher order in $U/c$  ($L'$ and $L''$ are finite there) and we
are left with the standard non-relativistic equation for all
$L(U^2/c^2)$:
 \beqray \deriv{\vvs}{t}=-\grad \chi, \label{nuuimat} \eeqray
where $\chi= -v^2/2$ can be identified as the Newtonian
gravitational potential. This holds when the gravitational field
is weak ($v\ll c$) and the motions are slow ($v_*\ll c$), but also
when only $U\ll c$. This can also be gotten directly from the
action, which for $U\ll c$ is $\int U^2dt$ up to a constant. This
limiting behavior is common to all theories with an $L(\c)$
Lagrangian including the standard acoustic line element.

Because the background flow is assumed to have a constant density
we see from the Bernoulli equation that $\chi$ equals the external
potential $\theta$ used to establish the background flow. (This
might point the way to introduce dynamics for the acoustic metric
through that of the external potential $\theta$.)
\par
Consider now the null geodesics of the acoustic metric
characterized by $U=c$, or $d\tau=0$. In theories for which
$L''/L'$ diverges when $U^2/c^2\rar 1$ the equation of motion
implies that $d(U^2)/dt=0$ if initially $U=c$, so $U$ remains
constant at this value. In other words, in such theories the null
world lines of the acoustic metric are solutions of the equation
of motion (this can be shown to hold even in flows with variable
background density). This is the case for the acoustic Lagrangian
itself, but also for our Lagrangians when $D\le 7$. For $D>7$,
$L,~L',~L''$ are finite as $\c\rar\infty$; for $5<D<7$, $L,~L'$
are still finite, but $L''$ diverges like $\c\^{(7-D)}$ (for $D=7$
$L''$ diverges logarithmically) and so does $L''/L'$; for $D<5$,
$L'$ diverges like $\c\^{(5-D)}$ (logarithmically for $D=5$) and
$L''$ still as $\c\^{(7-D)}$, so $L''/L'$ behaves as $\c^2$.
\par
Note that  $\vU\equiv 0$ is a solution (i.e., the body just moving
with the fluid). This is also a geodesic of the acoustic metric.
\par
We saw above several instances of solutions of our field equations
that are geodesics of the acoustic metric, and we shall see
another in the next subsection. But in general the solutions of
the Euler-Lagrange equations are not geodesics of the that metric,
as the particle action is not its arc length. It may be useful,
however, to generalize the acoustic proper-time interval and use
our action to define a Finslerian one (defined only for intervals
that are time- and null-like with respect to the acoustic metric)
whose geodesics are the solutions of our Euler-Lagrange equations.
For $D\le 3$, where the Lagranian diverges for $\c\rar \infty$
this is well defined only for time-like elements, but for $D>3$,
$d\hat\tau\propto [\ell(U^2/c^2)-\ell(1)]dt$ is well defined also
for null intervals of the acoustic metric (for which it vanishes).
Furthermore we saw that for $D<7$ this scheme embraces both
massive and massless particles.
\par
I demonstrate this for the more interesting case $D=4$: Subtract
from $L$ a constant equal to its relativistic limit:
 \beqray \hat L\equiv L-4mc^2=-Mc^2\c^{-1}\l(\c), \label{miopqa}\eeqray
 where $M\equiv 4m,$ and
 \beqray \l(\c)=\c/(1+\c). \label{butop}\eeqray
( When the background fluid density is not taken as a constant the
term we subtract in eq.(\ref{miopqa}) is position dependent, we
can then consider it as an external potential for the particle,
which does not couple to the fluid.)
 We can
then define a Finslerian line element
 \beqray  d\hat\tau(d\vec x,dt) =\c^{-1}\l(\c)dt, \label{nulka}\eeqray
defined for time- and light-like intervals, where here $\c$ stands
for $\{1-[d\vec x/dt-\vv(\vr)]^2\}^{-1/2}$. Clearly,
$d\hat\tau(d\vec x,dt)$ is homogeneous of order one as required.
Our particles follow geodesics of this Finslerian metric since the
action is $S=-Mc^2\int d\hat\tau$, and furthermore, phonons follow
its null geodesics, since $\d\hat\tau=0\Leftrightarrow d\tau=0,$
where $d\tau\propto \c^{-1}dt$ is the standard acoustic line
element, and we saw above that these too extremize the Finslerian
arc length (because they solve the equation of motion).

\subsection{Circular orbits in spherically symmetric space
times}
\par
Consider now circular orbit in a  spherically symmetric, or
axi-symmetric, configuration. Since $U^2$ is constant we can write
from eq.(\ref{cyuimat})
 \beqray \deriv{\vvs}{t}=\oot\grad v^2, \label{juliata}\eeqray
identical to the non-relativistic equation of motion. These orbits
are thus geodesics for all choices of $L(U^2/c^2)$ including the
acoustic one and all of our Lagrangians. For a circular orbit
$\deriv{\vvs}{t}=-v_*^2r^{-2}\vr$, and  $v^2$ is a function of $r$
($\vv$ is not necessarily radial). So the relation between the
velocity and the radius is:

\beqray v_*(r)=v(r)f^{1/2}, \label{uytr} \eeqray

where $v=|\vv|$, $f\equiv -\deriv{ln~v}{ln~r}$.
 Because I assumed a constant
density the continuity equation and zero vorticity condition
dictates that $f$ is determined by the symmetry of the flow. For
example, in purely radial flow in $D$ space dimensions $f=D-1$.
\par
When $\vv$ is radial $|\vec U|=v(1+f)^{1/2}$, so subsonicity
dictates for massive particles that $v<c(1+f)^{-1/2}$, which sets
the limit for the innermost circular orbit. Equality corresponds
to a ``photon'' orbit. In the canonical acoustic black hole
configuration (e.g. \cite{visser98}), where $v=c(r_h/r)^f$ with
$r_h$ the horizon radius, this implies $r>r_h(1+f)^{1/2f}$ for
massive particles. This is analogous to to the radius occurring at
$3m=1.5 r_h$ for real Schwarzschild black holes in $3+1$
dimensions.
\par
In a vortex geometry in (2+1) dimensions
 \beqray\vv/c={-r_h\vr/r+(r_e^2-r_h^2)^{1/2}\vec e_\theta\over r}
\label{uytgf}\eeqray
 (e.g. \cite{visser98}) where $r_e$ is the radius of the ergosphere, $r_h$ is
that of the event horizon and $e_\theta$ is a unit vector in the
azimuthal direction (I take an ingoing flow but the results below
are the same for an outgoing one); so $f=1$. The minimum radius of
a circular orbit is

\beqray r_L^\mp=[2r_e(r_e\mp \sqrt{r_e^2-r_h^2}]^{1/2},
\label{nuia} \eeqray where $r_L^-$ is for prograde motion and
$r_L^+$ for retrograde one.

As in the real world $r_e$ is a static limit since $\vvs=0$ is not
permitted below it lest $\vU$ become supersonic. Also for $r<r_h$
$\vvs$ must have a component in the radial direction of the flow.
\par
I haven't fully checked the question of stability of the circular
orbits. But note that in our constant density configurations the
Newtonian potential $\chi$ is a power law of the radius:
$\chi\propto r^{-2}$ in the $D=2$ case and with a higher power in
higher dimensions. This means that the effective radial potential
(including the centrifugal barrier) for a particle with given
angular momentum has only a maximum for $D>2$. So there aren't any
stable bound orbits in the nonrelativistic case and the circular
orbits are unstable. For $D=2$, depending on the value of the
angular momentum an orbit is either unbound or goes through the
origin. This is not necessarily so in configurations with
non-constant background densities, but their discussion is beyond
what I wish to consider here.

\section{Discussion}
Evidently, flow models can provide quasi-realistic analogs for
relativistic inertia of massive particles and their behavior in
gravitational fields. I am presenting these models in the twofold
hope that they can inspire us in understanding the origins and the
validity limits of genuine inertia, and that by considering how
these models respond to tweaking we can learn about possible
modification of standard physics in the real world. For example,
looking where and how our approximations break down we can gain
insight as to where and how standard dynamics may go awry. Such
departures may include breakdown of standard dynamics at high
$\c$, time non-locality of the particle action, and fluid-mediated
interactions between particles, all of which are not part of
standard dynamics. Such models can also be used to enlighten us on
how local dynamics might be affected by cosmology at large.
Cosmological expansion may be included in the context of fluid
analogs and could model, for instance, cosmological variations of
particle masses (through variations in the fluid density, which
enters the normalization of the masses), or variations in the
speed of light. My hope in this connection is to simulate the
dynamics implied by MOND, which revolves around an acceleration
constant, $a_0$ that turns out to be of the order of the cosmic
acceleration.
\par
The fact that the more realistic models emerge for higher space
dimensions than we seem to be living in is not disconcerting.
Recent work on membrane universes has taught us that while most of
the physical objects we deal with may be confined to sub-manifold
of lower dimensions some aspects of physics, such as gravity, may
be probing the higher dimensional aspects of space-time. In our
models we could, for example, envisage particles moving in a fluid
in $D$ dimensions space, but somehow confined to reside in a
three-dimensional sub-manifold. This would give rise to
$D$-dimension inertia in a lower dimensional effective space.
\par
My main purpose in this paper is to demonstrate the concept:
Instead of considering weak perturbations that are part of the
background itself, and which thus move with a speed dictated by
the characteristic speeds of the background, define the particle
as an externally dictated perturbation that breaks the field
equation of the background, but that can otherwise move freely on
the background field. Various extensions and generalizations
suggest themselves that are worth exploring. For example, we can
define other types of particles, or permit non-rigid particles
with responsive intrinsic structure. This would produce longer
range interactions between particles similar in nature to van der
Vals interactions between neutral charge distributions (our rigid
particles interact only on contact). Such particles with
dynamical, internal degrees of freedom; e.g., with the different
charges in a source system connected by ``springs'', may also
serve as detectors for (phononic) Unruh radiation, as the internal
degrees of freedom will couple to the phononic field.
\par
And, we can generalize this idea to the whole gamut of analog
models for which photon propagation can be simulated (e.g.,
\cite{barcelo05}). One possibility, for example, is to look at
small charge distributions (of vanishing total charge) in the
context of non-linear electrodynamics where the action of the
electromagnetic field is not the invariant $F_{\m\n}F^{\m\n}$ but
some function of it.
\par
Finally note that as things now stand, our particles cannot serve
as sources for a mock gravitational field through their effect on
the fluid: Their mass $m$ is not an active gravitational mass. So,
they do not help towards constructing an analog of the
Einstein-Hilbert action.

\appendix
\section{Calculation of the Lagrangian for a source particle}
Here I calculate the energy integral in eq.(\ref{kp}) for a
spherical distribution of sources $\hs(r)$. Divide the
distribution into concentric thin shells of radii $r_i$ and total
charges $q_i$. The integral is twice the electrostatic energy of
the system made of these shells all stretched by a factor $\c$ in
the $z$ direction. The stretching is only of the geometry without
thinning the density. Each spherical shell becomes a homoeoid: a
shell bound by two concentric, oriented ellipsoids of the same
axes ratio $\c$, with the original density inside; so, the
resulting homoeoid $i$ has a total charge $Q_i=\c q_i$. Write now
the energy as
 \beqray E=\sum_i E_i +\sum_{i< j}E_{ij},
\label{bein} \eeqray
 where $E_i$ is the self energy of shell $i$ and
$E_{ij}$ is the interaction energy betweens shells $i$ and $j$
($i$ is interior to $j$). A homoeoid produces a constant potential
inside its cavity ($\varphi_i$ for homoeoid $i$) and if the
homoeoid is thin, as here, this is also the potential on the
shell. Thus, $E_i=Q_i\varphi_i/2$ and $E_{ij}=Q_i\varphi_j$. We
can most easily calculate $\varphi_i$ as the value of the
potential at the center of the cavity and this is simply (for
$D>2$)
 \beqray
\varphi_i(\c)&=&\varphi_i(1)\c{\ODo\over\OD}\int_0^\pi
{sin^{D-2}\theta~d\theta\over
[1+(\c^2-1)cos^2\theta]^{(D-2)/2}}\nonumber\\&=&
\varphi_i(1){\ODo\over\OD}2\c^{3-D}\times\nonumber\\&&\int_0^{\pi/2}
{sin^{D-2}\theta~d\theta\over
[1-(U/c)^2sin^2\theta]^{(D-2)/2}},\label{byuta}\eeqray
 where $\OD$
is the $D$-dimensional solid angle ($\OD=\ODo \int_0^\pi
sin^{D-2}\theta d\theta.$), and $\varphi_i(1)=q_i/r_i\^{D-2}$ is
the potential for the unstretched shell. The integral can be
expressed using a Gauss hypergeometric function (using formula
3.681.1 in \cite{gr45}, henceforth GR):
 \beqray \int_0^{\pi/2}=\oot B({D-1\over
2},\oot)F({D-2\over 2},{D-1\over 2};{D\over 2};{U^2\over
c^2}).\label{juosa} \eeqray It can be shown that
$B[(D-1)/2,1/2]=\OD/\ODo$. Also use formula 9.131.1 in GR to
further simplify and get
 \beqray E=E_0\c F(1,\oot;{D\over 2};{U^2\over c^2}), \label{nuhya} \eeqray
 so for the Lagrangian one has
 \beqray L=L_0F(1,\oot;{D\over 2};{U^2\over c^2}). \label{nujuo} \eeqray

\par
For $D=2$ one finds that the dependence of the energy on $\c$ is
of the form $E=\c^2 E_0+\c^2 Q^2f(U)$, where $Q$ is the total
charge and
$f(U)=(1/\pi)\int_0^\pi~d\theta~ln[1+(\c^2-1)cos^2\theta]$. Since
we have to take a vanishing total charge, we are left with $E=\c^2
E_0$, so $L=L_0 \c$, which also conforms with eq.(\ref{nujuo})
since $F(1,\oot;1;{U^2\over c^2})=\c$. Calculation for $D=1$ is
also straightforward and the result is also given by
eq.(\ref{nujuo}): $L=L_0\c^2$.

\section{Calculation of the Lagrangian for a potential particle}
\par
Here I calculate the Lagrangian for a spherical potential
particle. We need the middle term in eq.(\ref{pointo}), which can
be written as
 \beqray
-{\r\over 2\c}\int\drD'\heta(\vr')q(\vr')= {\r\over
2\c}\int\drD'[\grad\heta(\vr')]^2,\label{mubtara}\eeqray
  with
$q(\vr')=-c^{-2}\c\vU\cdot\grad'\hpsi[\vr(\vr')]$
 and
$\heta$ that solves $\Delta'\heta=q(\vr')$. The integral is twice
the electrostatic energy of the charge distribution $q(\vr')$,
which is produced by a polarized body with unidirectional
polarization $\vec P(\vr')=-c^{-2}\c\hpsi[\vr(\vr')]\vU$. We start
from a spherical body having some $\hpsi(r)$ and divide it into
thin spherical shells of radii $r_i$ and thicknesses $dr_i$ of
constant $\hpsi(r_i)$. These are stretched into thin, concentric,
nested homoeoids of minor axes $r_i$ and axes ratio $\c$ with
constant polarization $P(r_i)$ along the major axis. We need the
energy of this configuration. It is well known that the field
inside an ellipsoid with uniform polarization along the major
axis, is constant and proportional to the polarization with the
proportionality factor depending only on the axes ratio. In our
case, for a constant $\hpsi$ we write
$\grad'\heta=\emph{d}(\c)\vec P=-\emph{d}(\c)\c c^{-2}\hpsi\vU$.
(In the context of magnetostatics $\emph{d}(\c)$ is called the
demagnetizing factor.) This means that a thin homoeoid with
uniform polarization $\vec P$ along its major axis produces a
vanishing field inside it, and thus the interaction energy of two
nested homoeoids such as ours vanishes. The Lagrangian produced by
our stretched configuration is then the sum of the contributions
of the self energies of the individual thin homoeoids. Each thin
homoeoid may be viewed as an infinitesimally thin dipole bilayer
whose different elements thus do not interact with each other. Its
energy is then the surface integral of the energy of a charged
parallel-plate capacitor $E= (1/2)\int ~ \d a \sigma^2 dS,$ where
sigma is the surface density of the charge, and $\d a$ is the
thickness, both dependent on the polar angle $\theta'$ (relative
to the major axis) at the position of the integration point $q'$
on the homoeoid. Use instead as variable the polar angle $\theta$
at point $q$ on the spherical shell which was stretched into $q'$;
we then have $\sigma=(U\c/c^2)\hpsi cos\theta(cos^2\theta+\c^2
sin^2\theta)^{-1/2},$
 $\d a=\c dr_i (cos^2\theta+\c^2 sin^2\theta)^{-1/2}$,
 $dS=(cos^2\theta+\c^2
 sin^2\theta)^{1/2}\ODo r_i^{D-1}sin^{D-2}\theta d\theta,$
 where $\OD$ is the $D$-dimensional solid angle $\OD=\ODo
\int_0^\pi sin^{D-2}\theta d\theta.$

Thus the required contribution to $L$ of the $i$th shell (for
$D\ge 2$) can be written as
 \beqray dL_i&=&\oot{\r\over
c^4}U^2\c^2\hpsi^2(r_i)\OD r_i^{D-1}dr_i{\ODo\over
 \OD}\times\nonumber\\&&\int_0^\pi {cos^2\theta sin^{D-2}\theta \over
 cos^2\theta+\c^2 sin^2 \theta}d\theta.
\label{vuopa} \eeqray
 Summing over the shells we replace
$\hpsi^2(r_i)\OD r_i^{D-1}dr_i\rar \int\hpsi^2 \drD$

 The integral over $\theta$ is
$$2\int_0^{\pi/2} {cos^2\theta sin^{D-2}\theta \over
1-(1-\c^2) sin^2 \theta}d\theta, $$
  which can be read off formula 3.681.1 in GR to be

$${\Gamma[(D-1)/2]\Gamma(3/2)\over \Gamma[(D+2)/2]}F(1,{D-1\over
2};{D+2\over 2};1-\c^2). $$
 The factor in front can be shown to
give $\OD/D\ODo$. I also use formula 9.131.1 in GR to transform
$$F(1,{D-1\over 2};{D+2\over 2};1-\c^2)=\c^{-2}F(1,{3\over
2};{D+2\over 2};{U^2\over
 c^2}),$$
 and then 9.137.12 in
GR to write
 $$ F(1,{3\over 2};{D+2\over 2};{U^2\over
 c^2})=D\left({U^2\over
 c^2}\right)^{-1}[F(1,\oot;{D\over 2};{U^2\over
 c^2})-1],$$
 then adding the last term in eq.(\ref{pointo}) I get finally
 \beqray L=L_0F(1,\oot;{D\over 2};{U^2\over
 c^2}),\label{vuuta} \eeqray
with
 \beqray L_0\equiv {\r\over 2c^2}\int\hpsi^2 \drD.\label{jiuter}\eeqray
  For $D=1$ it is straightforward to solve directly for
$\eta$ from eq.(\ref{mseui}) and substitute in eq.(\ref{pointo}).
It turns out that eq.(\ref{vuuta}) is still valid  giving
$L=L_0\c^2$. The same is true for $D=2$ where we have $L=L_0\c$.
\par
As a byproduct of the above calculation we get the expression for
the $D$ dimensional demagnetizing factor for a prolate ellipsoid
magnetized along the symmetry axis. Consider a case where $\hpsi$
is constant inside the stretched ellipsoid in which
$\grad'\heta=-\emph{d}(\c)c^{-2}\c\hpsi\vU$. What I calculated
above is the quantity
 \beqray
&&{\r\over 2c^2}\int\drD~\heta(\vU\cdot\grad\hpsi)= -{\r\over
2c^2}\int\drD~\hpsi(\vU\cdot\grad\heta)\nonumber\\&=& -{\c\r\over
2c^2}\int\drD~\hpsi(\vU\cdot\grad'\heta)
\nonumber\\&=&\emph{d}(\c)(\c^2\r/2c^2)(U/c)^2\int\hpsi^2\drD.\eeqray
 This equals $(\r/2c^2)(\int\hpsi^2 \drD) [F(1,\oot;{D\over
2};{U^2\over c^2})-1]$, as we found above. Comparing the two
expressions one gets
 \beqray \emph{d}(\c)=(\c^2-1)^{-1}[F(1,\oot;{D\over 2};{U^2\over c^2})-1].\label{butra}\eeqray
For $D=3$ this gives
 \beqray &&d(\c)={1\over \c^2-1}\times\nonumber\\&&\left[{\c\over 2
(\c^2-1)^{1/2}}ln\left({\c+(\c^2-1)^{1/2}\over\c-(\c^2-1)^{1/2}}\right)
-1\right],\label{musasa}\eeqray
 which reproduces the result found in \cite{osborn45}.

\section{The rest mass of a source particle}
The rest energy of a source particle is the energy difference
between two configurations, one of a uniform fluid at rest, the
other likewise but with the source inserted. There are subtleties
involved in the determination of this difference as the two
configurations have infinite mass and energy. I adopt the
following scheme: Consider a container of finite volume $V$ much
larger than that of the particle and filled with a static
homogeneous fluid at the reference density $\r$. Consider now a
spherical source of vanishing total out-flux somewhere inside the
container, in a steady state. Write the source density as
$\hs(\vr)=\r\Delta\Phi=\r\div\vu$ for some $\vu=\grad\Phi$; and if
the support of $\hs$ is within radius $R_0$ of its center we
deduce from Gauss theorem that $\Phi$ and $\vu$ vanish everywhere
outside $R_0$. Writing the density as $\r+\z$, the continuity
equation is $\div[(\r+\z)\vv-\r\vu]=0$, with $\vv$ and $\vu$
radial from the center of the source; hence
 \beqray \vv=(1+\z/\r)^{-1}\vu \label{lululu}\eeqray
is exact and $\vv$ vanishes everywhere outside the source. The
Bernoulli equation is
 \beqray \oot v^2+h(\r+\z)=constant\equiv
 h(\tilde\r).\label{kukuku}\eeqray
Since outside the source $\vv=0$ it follows that the density is
constant there and equals $\tilde\r$. I require this configuration
with the source to have the same total fluid mass as the reference
configuration, so $\int_V \z\drD=0$ and this closes the set of
algebraic equations that determines the configuration completely.
The energy of the source configuration relative the reference one
is then
 \beqray E_r=\int_V[\oot(\r+\z) v^2+e(\r+\z)-e(\r)]\drD.\label{mumumu}\eeqray
It is evident that neither the run of $\z$ and $\vv$ inside the
source nor the value of $\tilde\r$, the constant density outside
the source, depend on the position of the source inside the
container, nor on the shape of the container (though its volume
does enter). So far everything is exact. When the source may be
considered a weak perturbation as in our case ($\z\ll\r$, $v\ll
c$) we can expand to lowest order in $\z$: Bernoulli's equation
gives
 \beqray \z(\vr)\approx\tilde\z-{\r v^2(\vr)\over 2 c^2}, \label{kuluku}\eeqray
where $\tilde\z\equiv\tilde\r-\r$; the continuity equation gives
$\vv\approx\vu\approx\grad\Phi$; and the preserved-total-mass
constraint gives
 \beqray \tilde\z\approx{\r\over 2c^2V}\int_S u^2\drD, \label{muku}\eeqray
where the integral is over the volume of the source. Clearly
$\tilde\z$ vanishes in the limit of infinite $V$. Writing
$e(\r+\z)-e(\r)\approx h(\r)\z$, the fact that $\z$ integrates to
zero means that there is no first order contribution to the
intrinsic energy difference. The second order contribution $\int
\oot {c^2\over \r}\z^2\drD$ is higher order in $u^2/c^2$. We are
left with the dominant contribution
 \beqray E_r &\approx& \oot\int_S\r (\grad\Phi)^2\drD=-\oot\int_S\Phi\hs\drD
 \nonumber\\&=&-\oot\int_S\heta\hs\drD=L_0,\label{mumaku}\eeqray
which depends neither on the position of the source nor on the
volume of the container or its shape, and we can identify it as
the rest energy for large $V$. It is seen that $E_r=E_0/\r\OD$, so
we finally get
  \beqray m_0={D\over 2}m. \label{nunumu}\eeqray
One might also worry about the elastic energy of the container's
wall, which is different in the reference configuration--where the
wall is subject to pressure $p(\r)$--and the one with the source
where the pressure on the wall is $p(\tilde\r)\approx
p(\r)+c^2\tilde\z$. But this energy scales with the size of the
container as $A\tilde\z\propto V^{-1/3}$ ($A$ the area of the
wall), so it can be neglected for large volumes.
\par
When more than one particle is present without overlap they do not
interact and the total energy is the sum of the rest masses. (Due
to the non-linearity of the energy there is interaction when
particles overlap, but since $\Phi=0$ outside particles there is
no long range interaction--similar to the case of two rigid
spherical charge distributions, each of vanishing total charge.)


\end{document}